\documentclass[aps,pra,reprint,superscriptaddress,nofootinbib]{revtex4-2}
\usepackage[T1]{fontenc}
\usepackage[utf8]{inputenc}
\usepackage{amsmath,amssymb,bm,graphicx}
\usepackage[hidelinks]{hyperref}
\usepackage{microtype}

\begin{document}
\title{Bell-certified retardance polarimetry from CHSH correlators}
\author{J. Sumaya-Martinez}
\email{jsm@uaemex.mx}
\affiliation{Faculty of Sciences, Universidad Autónoma del Estado de México, Toluca 50000, México}
\date{\today}

\begin{abstract}
Bell--CHSH measurements with polarization-entangled photons are usually reported only as nonlocality witnesses. Here we show that, for retardance sensing, the same coincidence data can also provide a quantitative metrological certificate. For a local birefringent phase $\phi$ encoded on one photon, we derive the classical Fisher information directly from experimentally accessible CHSH correlators $E_{xy}(\phi)$ and obtain an explicit lower bound in terms of the measured correlators and the phase slope $S'(\phi)$ of the Bell parameter. In the unbiased-marginal two-qubit regime, this leads to a dual-use result: the canonical CHSH analyzer settings are simultaneously Bell-optimal and Fisher-optimal, saturating the quantum Fisher information for retardance estimation. We further show that the certificate remains informative under visibility loss, finite counts, and moderate departure from the Bell-violation threshold, so useful retardance sensitivity can persist even when nonlocality certification is weak or absent. Relative to tomography-based entangled-photon polarimetry, the method converts a standard Bell-calibration run into a sensitivity statement with rigorous lower confidence bounds using only coincidence correlators. The scheme is implementable with standard SPDC sources, wave plates, and polarizing beam splitters.
\end{abstract}
\maketitle

\section{Introduction}
Bell inequalities quantify incompatibility between observed correlations and local-hidden-variable models, while quantum metrology asks how precisely a parameter encoded in a quantum state can be estimated. Its fundamental benchmark is the quantum Fisher information (QFI), which upper-bounds the classical Fisher information (FI) obtainable from any measurement \cite{Pezze2009,Hyllus2012,Toth2012,Pezze2018}. Bell nonlocality and QFI are not equivalent resources in general \cite{Frowis2019}. This motivates an operational question: when a polarization Bell test is already being performed, can the same data be converted into a rigorous metrological certificate without tomography or a separate estimator-specific calibration?

Here we answer this question for single-parameter retardance polarimetry with entangled photons. We assume the standard CHSH coincidence data at two settings per side together with a known local phase encoding on one photon. In the unbiased-marginal regime, we derive an exact correlator-level FI expression and a compact CHSH-to-precision inequality. In the ideal two-qubit model the canonical CHSH settings are a dual optimum: they maximize Bell violation at $\phi=0$ and saturate the QFI for retardance estimation across the full phase range.

The contribution is not a universal equivalence between Bell violation and metrological usefulness. Useful FI can survive below the Bell-violation threshold. Rather, the same Bell-calibration data can certify a nontrivial lower bound on retardance sensitivity, and finite-count confidence bounds can be attached directly to that certificate.

\section{Model and assumptions}
\subsection{Trust model}
Our ``Bell-certified'' statements are semi-device-independent. We assume: (i) a two-outcome measurement model with well-defined settings $x,y$; (ii) a known local encoding $U_A(\phi)$, or more generally a known channel family; and (iii) i.i.d. trials at fixed $\phi$ and settings. The simple correlator formula below additionally assumes unbiased single-party marginals. With biased marginals the joint distribution is
\begin{equation}
P(a,b|x,y,\phi)=\frac14[1+a\,m_x(\phi)+b\,n_y(\phi)+abE_{xy}(\phi)],
\end{equation}
and the FI should be evaluated from the full probabilities.

\section{Retardance encoding and CHSH measurements}
Let $H\equiv|0\rangle$ and $V\equiv|1\rangle$, and take
\begin{equation}
|\Phi\rangle=\frac{|HH\rangle+|VV\rangle}{\sqrt2}.
\end{equation}
A local birefringent element on Alice implements
\begin{equation}
U_A(\phi)=e^{-i\phi\sigma_z/2},\qquad
|\psi_\phi\rangle=(U_A(\phi)\otimes I)|\Phi\rangle.
\end{equation}
Alice and Bob choose $x,y\in\{0,1\}$ and record outcomes $a,b\in\{\pm1\}$. The correlators are
\begin{equation}
E_{xy}(\phi)=\sum_{a,b=\pm1}ab\,P(a,b|x,y,\phi)
=\langle A_x\otimes B_y\rangle_{\rho_\phi},
\end{equation}
and the CHSH parameter is
\begin{equation}
S(\phi)=E_{00}+E_{01}+E_{10}-E_{11}.
\end{equation}

\section{Fisher information from CHSH data}
For unbiased marginals,
\begin{equation}
P(a,b|x,y,\phi)=\frac{1+abE_{xy}(\phi)}4.
\end{equation}
If $p_{xy}$ is the sampling distribution of analyzer settings, the FI is
\begin{equation}
F_C(\phi)=\sum_{xy}p_{xy}\sum_{ab}P(a,b|x,y,\phi)
[\partial_\phi\ln P(a,b|x,y,\phi)]^2.
\end{equation}
Direct substitution gives
\begin{equation}
\boxed{
F_C(\phi)=\sum_{xy}p_{xy}
\frac{[\partial_\phi E_{xy}(\phi)]^2}{1-E_{xy}^2(\phi)}.}
\end{equation}

Define $c_{00}=c_{01}=c_{10}=1$, $c_{11}=-1$ and
$S'(\phi)=\sum_{xy}c_{xy}\partial_\phi E_{xy}$. Cauchy--Schwarz yields
\begin{equation}
\boxed{
F_C(\phi)\ge
\frac{[S'(\phi)]^2}
{\displaystyle\sum_{xy}\frac{1-E_{xy}^2(\phi)}{p_{xy}}}.}
\end{equation}
For uniform sampling, $p_{xy}=1/4$,
\begin{equation}
F_C(\phi)\ge
\frac{[S'(\phi)]^2}
{4\sum_{xy}[1-E_{xy}^2(\phi)]}.
\end{equation}
This is the CHSH-to-precision certificate.

\section{Dual-use optimality}
For equatorial polarization observables
\begin{equation}
\sigma_\theta=\cos\theta\,\sigma_x+\sin\theta\,\sigma_y
\end{equation}
with $A_x=\sigma_{\alpha_x}$ and $B_y=\sigma_{\beta_y}$, one obtains
\begin{equation}
E_{xy}(\phi)=\cos(\alpha_x-\beta_y+\phi).
\end{equation}
Choose the canonical CHSH angles
\begin{equation}
\alpha_0=0,\quad \alpha_1=\frac\pi2,\quad
\beta_0=\frac\pi4,\quad \beta_1=-\frac\pi4.
\end{equation}
Then
\begin{equation}
S(\phi)=2\sqrt2\cos\phi,\qquad S'(\phi)=-2\sqrt2\sin\phi.
\end{equation}
For every setting pair the FI ratio equals unity; hence uniform sampling gives
\begin{equation}
F_C(\phi)=1.
\end{equation}
For the unitary generator $G=\sigma_z/2$, the Bell state has
\begin{equation}
F_Q=4\,{\rm Var}(G)=1,
\end{equation}
so
\begin{equation}
\boxed{F_C(\phi)=F_Q(\phi)=1.}
\end{equation}
The canonical CHSH settings therefore maximize the Bell signal at $\phi=0$ while saturating the QFI for retardance estimation.

\begin{figure}[t]
\includegraphics[width=\columnwidth]{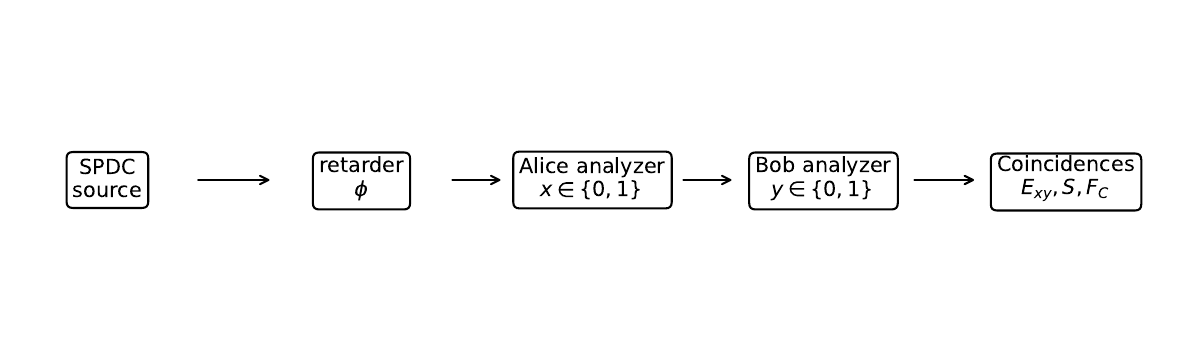}
\caption{Conceptual dual-use scheme. The same coincidence stream provides the correlators $E_{xy}(\phi)$, the CHSH parameter $S(\phi)$, and a Fisher-information sensitivity certificate.}
\end{figure}

\section{Robustness}
A depolarizing visibility model is
\begin{equation}
\rho_v(\phi)=v|\psi_\phi\rangle\langle\psi_\phi|+\frac{1-v}{4}I,
\end{equation}
which gives
\begin{equation}
E_{xy}(\phi)=v\cos(\alpha_x-\beta_y+\phi),\qquad
S(\phi)=2\sqrt2\,v\cos\phi.
\end{equation}
The FI becomes
\begin{equation}
F_C(\phi)=\sum_{xy}p_{xy}
\frac{v^2\sin^2(\alpha_x-\beta_y+\phi)}
{1-v^2\cos^2(\alpha_x-\beta_y+\phi)}.
\end{equation}
Bell violation requires $v>1/\sqrt2$ at the optimal point, but the FI remains positive for any $v>0$. Thus Bell violation is not itself the metrological resource.

Phase diffusion $\phi\rightarrow\phi+\xi$, $\xi\sim\mathcal N(0,\sigma^2)$, gives an effective coherence factor $e^{-\sigma^2/2}$,
\begin{equation}
E_\xi[\cos(\theta+\phi+\xi)]
=e^{-\sigma^2/2}\cos(\theta+\phi).
\end{equation}
Small analyzer errors enter as $\alpha_x\to\alpha_x+\Delta\alpha_x$ and
$\beta_y\to\beta_y+\Delta\beta_y$ and can be propagated through the same observable correlators.

\begin{figure}[t]
\includegraphics[width=\columnwidth]{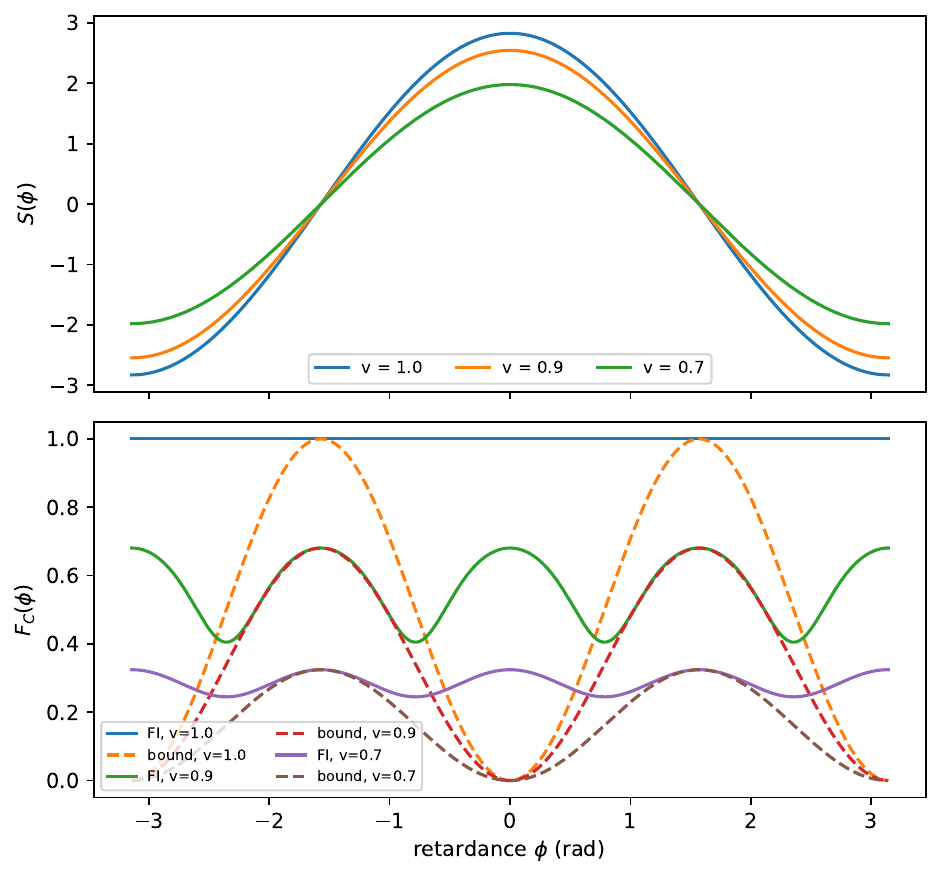}
\caption{Ideal-model behavior for canonical CHSH settings under visibility loss. Top: $S(\phi)$. Bottom: classical FI (solid) and the CHSH-based lower bound (dashed). For $v=1$, $F_C=F_Q=1$.}
\end{figure}

\section{Relation to conventional polarimetry}
Equation (8) has the familiar slope-squared-over-variance structure. Each analyzer pair contributes a parameter-response slope, while $1-E_{xy}^2$ is the dichotomic output variance. The CHSH framing adds a certification layer: the same dataset carries both phase-response information and a nonclassicality witness. Unlike tomography-based approaches \cite{Pedram2024}, the lower bound can be reported directly from Bell correlators and a controlled phase modulation.

\section{Finite-data certification}
Let $N^{xy}_{ab}$ denote coincidence counts and
$N_{xy}=\sum_{ab}N^{xy}_{ab}$. The estimator
\begin{equation}
\widehat E_{xy}=\frac{N_{++}+N_{--}-N_{+-}-N_{-+}}{N_{xy}}
\end{equation}
is the sample mean of $t=ab\in\{\pm1\}$. Its variance is
$(1-E_{xy}^2)/N_{xy}$.

For a distribution-free confidence interval, Hoeffding's inequality \cite{Hoeffding1963} gives
\begin{equation}
\Pr(|\widehat E_{xy}-E_{xy}|\ge\epsilon_{xy})\le\alpha,\qquad
\epsilon_{xy}=\sqrt{\frac{2}{N_{xy}}\ln\frac{2}{\alpha}}.
\end{equation}
Therefore
\begin{equation}
|E_{xy}|\ge \underline{|E_{xy}|}
=\max\{0,|\widehat E_{xy}|-\epsilon_{xy}\}.
\end{equation}

Estimate the derivative by a symmetric modulation:
\begin{equation}
\widehat S'=\frac{\widehat S(\phi_0+\delta)-\widehat S(\phi_0-\delta)}{2\delta}.
\end{equation}
If $\Delta S_\pm=\sum_{xy}\epsilon^\pm_{xy}$, then the statistical derivative radius is
\begin{equation}
\Delta S'=\frac{\Delta S_++\Delta S_-}{2\delta}.
\end{equation}
For the cosine model, a conservative finite-difference truncation term is
\begin{equation}
\Delta_{\rm fd}=\frac{\delta^2}{6}\,2\sqrt2.
\end{equation}
Thus
\begin{equation}
\underline{|S'|}=
\max\{0,|\widehat S'|-\Delta S'-\Delta_{\rm fd}\}.
\end{equation}
A conservative lower confidence bound on the FI is then
\begin{equation}
\boxed{
F_C(\phi_0)\ge \underline F_C=
\frac{\underline{|S'|}^{\,2}}
{\displaystyle\sum_{xy}
\frac{1-\underline{|E_{xy}(\phi_0)|}^{\,2}}{p_{xy}}}.}
\end{equation}
For twelve correlators (four at $\phi_0$ and four at each of $\phi_0\pm\delta$), choosing per-correlator failure probability $\alpha=\eta/12$ gives joint confidence at least $1-\eta$ by a union bound. Empirical-Bernstein intervals can tighten the certificate \cite{Maurer2009}.

\begin{figure}[t]
\includegraphics[width=\columnwidth]{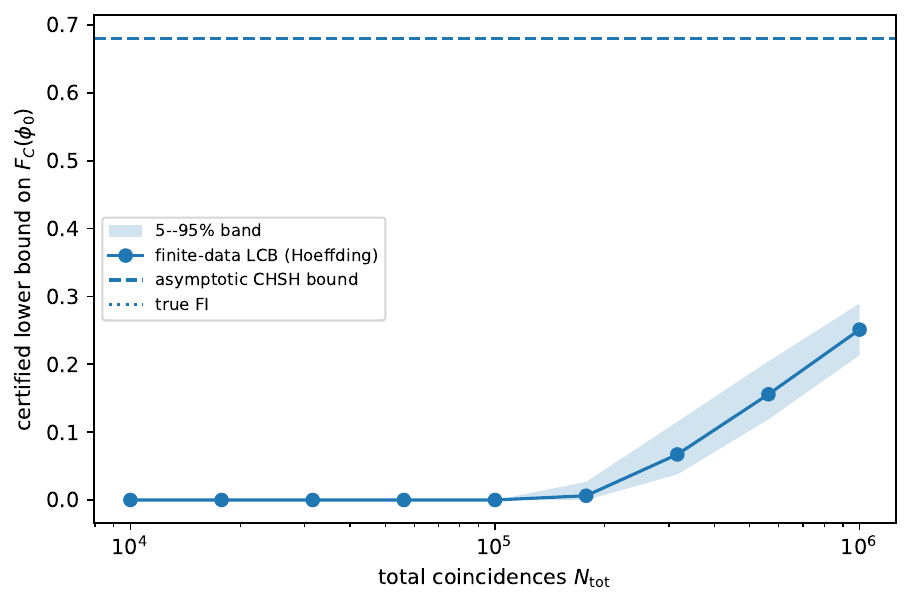}
\caption{Deterministic Monte Carlo illustration of finite-data certification at $\phi_0=\pi/2$, $v=0.9$, $\delta=0.05$ rad. The median Hoeffding lower-confidence bound approaches the asymptotic CHSH certificate as the coincidence budget increases. The included Python script reproduces this figure.}
\end{figure}

\section{Implementation}
A minimal experiment uses SPDC to generate polarization-entangled photon pairs, a birefringent sample or tunable retarder in Alice's arm, and HWP/QWP+PBS analyzers on both sides. From $N^{xy}_{ab}(\phi)$ one estimates $E_{xy}$ and $S$. A small calibrated modulation $\pm\delta$ provides $S'$. The same data stream then yields a CHSH test, the FI through the exact correlator formula, and a guaranteed lower bound through the CHSH-to-precision inequality.

\section{Discussion}
The central statement is narrow: in polarization-entangled retardance metrology, CHSH data can be promoted from a nonlocality witness to a quantitative sensitivity certificate. No density-matrix reconstruction is required for the correlator bound, and in the ideal model the same analyzer settings that maximize the Bell signal also saturate the QFI.

The result should not be interpreted as a universal equivalence between Bell violation and metrological advantage. Large QFI does not imply Bell correlations in general \cite{Frowis2019}, and useful FI can remain even when the CHSH inequality is not violated. The certification is also not loophole-free: the encoding model and setting labels are trusted, and the clean closed form assumes unbiased marginals. These assumptions define the scope of the result.

Natural extensions include a direct comparison against optimized separable polarimetry at equal photon budget, multiparameter polarization estimation using QFI matrices, robust certification under calibration drift and unequal detection efficiencies, and replacement of CHSH by steering or measurement-device-independent witnesses \cite{Bardyn2009,Kocsis2015}.

\section{Conclusion}
Standard Bell-test data can carry more than a nonlocality witness. For a known local retardance encoding, CHSH correlators provide an exact FI expression in the unbiased-marginal regime and an experimentally accessible lower bound on precision. Canonical CHSH analyzer settings saturate the QFI for the ideal Bell state, while finite-data confidence procedures allow the sensitivity statement to remain statistically controlled under realistic counts and visibility loss.

\appendix
\section{Derivation of the correlator FI}
For fixed $x,y$, write $E=E_{xy}(\phi)$ and
$P_{ab}=(1+abE)/4$. Then
\begin{align}
\sum_{ab}P_{ab}[\partial_\phi\ln P_{ab}]^2
&=\sum_{ab}\frac{(\partial_\phi P_{ab})^2}{P_{ab}}\\
&=(\partial_\phi E)^2\sum_{ab}\frac{1}{4(1+abE)}\\
&=\frac{(\partial_\phi E)^2}{1-E^2}.
\end{align}

\section{Proof of the CHSH-to-precision bound}
Define
\begin{equation}
w_{xy}=\sqrt{p_{xy}}\frac{\partial_\phi E_{xy}}{\sqrt{1-E_{xy}^2}},
\quad
d_{xy}=\frac{c_{xy}}{\sqrt{p_{xy}}}\sqrt{1-E_{xy}^2}.
\end{equation}
Then $F_C=\sum w_{xy}^2$ and $S'=\sum d_{xy}w_{xy}$. Cauchy--Schwarz gives
\begin{equation}
(S')^2\le
\left[\sum_{xy}\frac{1-E_{xy}^2}{p_{xy}}\right]F_C,
\end{equation}
which yields the stated lower bound.

\section{Biased marginals}
If $m_x$ or $n_y$ is appreciably nonzero, the full likelihood in Eq.~(1) should be used. One may explicitly include $m_x,n_y$ in the likelihood or enforce outcome symmetrization by random known label flips, at the cost of reduced visibility.

\section{Confidence-budget optimization}
The union bound permits nonuniform allocation of failure probabilities $\{\alpha_j\}$ as long as $\sum_j\alpha_j\le\eta$. For Hoeffding radii one can approximately equalize weighted radii within each phase group. If $w_j\epsilon_j=\kappa$, then
\begin{equation}
\alpha_j=2\exp\left[-c\,\frac{N_j}{w_j^2}\right],
\end{equation}
with $c>0$ chosen so that the allocated group budget is saturated. This is a one-dimensional monotone solve and can be done before data acquisition.

\section*{Funding}
Universidad Autónoma del Estado de México (UAEMex).

\section*{Acknowledgments}
The author thanks colleagues and collaborators for helpful discussions.

\section*{Data availability}
Data underlying the results are available from the corresponding author upon reasonable request.

\section*{Code availability}
The package includes \texttt{make\_figures.py}, which reproduces the numerical figures.

\section*{Disclosures}
The author declares no conflicts of interest.

\bibliographystyle{apsrev4-2}
\bibliography{refs}
\end{document}